\documentclass[useAMS,usenatbib]{mn2e}

\usepackage{graphicx}

\title[Discovery of a galactic wind in the central region of M100]{Discovery of a galactic wind in the central region of M100}
\author[]{J. Jim\'enez-Vicente$^{1}$\thanks{E-mail:
jjimenez@ugr.es}, A. Castillo-Morales
$^{2}$, E. Mediavilla$^{3}$ and E. Battaner$^{1}$\\
$^{1}$Dpto de F\'{\i}sica Te\'orica y del Cosmos. Universidad de Granada, Spain.\\
$^{2}$Dpto de Astrof\'{\i}sica y C.C. de la Atm\'osfera. Univ. Complutense de Madrid, Spain.\\
$^{3}$Instituto de Astrof\'{\i}sica de Canarias. Tenerife. Spain.}
\begin{document}



\maketitle
\label{firstpage}
\begin{abstract}
We report the discovery of a galactic wind in the central region of the 
galaxy M100.  This result is based on a careful 2D spectroscopic 
study performed on observations made with the fibre system INTEGRAL on the WHT.  
The primary 
evidence of the wind is the presence of blueshifted interstellar NaD absorption 
lines. The velocity field of the absorbers show a clear rotation pattern but globally 
blueshifted ($\sim$ -115 km/s) with respect to the systemic velocity of the galaxy.
  The emission lines also present a blueward component arising from the 
ionized gas phase of the galactic wind. The velocity field of the ionized gas wind component shows no evidences of rotation but exhibits a pattern that can be interpreted in terms of the projection of an outflowing cone or shell.  The wind component  has [NII]/H$\alpha$ ratios of about 1.8, typical of shock ionization.
The ionized component of the wind can be identified with an expanding shell of 
shocked gas, and the neutral component with disk gas entrained in the wind at the
interface of the expanding shell with the galactic ISM. 
The galactic wind seems to be driven uniquely by the nuclear starburst. 
Our analysis indicates that a non negligible fraction of the wind
material might escape to the IGM.
In this case, if the wind detected in M100 were representative of similar
phenomena in other
galaxies with low to moderate activity,
the current estimates of metal and  dust content of the IGM might be drastically 
underestimated.
\end{abstract}
\section{Introduction}
Galactic winds are one of the main mechanisms by which galaxies expel metals, dust,
cosmic rays, magnetic fields, etc,  
to the intergalactic medium. This feedback makes winds an important 
phenomenon in our global understanding of galaxies. These winds 
are usually associated
to galaxies with strong nuclear activity or with starbursts
~\cite[see][and references therein]{veill05} which are powerful enough to
feed the winds. But very little is known of this phenomenon in quiescent,
 inactive galaxies. In this paper, we explore this phenomenon 
in a relatively quiet galaxy.
M100 is a nearby (D=16.1 Mpc) galaxy with low nuclear activity
\citep[cataloged as transition LINER/HII by][]{ho97a} 
 and a moderate star formation rate of a few solar masses per year ~\citep[according to its IR luminosity of ${\rm L_{IR}}=10^{10.39} {\rm L_\odot}$, ][]{sanders03}.
Despite this apparent quiescence, we will show that this galaxy presents
an unexpected fast wind emanating from its nuclear region.

Winds have been studied in highly inclined galaxies where the 
extraplanar gas can be directly seen through imaging of the emission 
lines of the ionized 
gas ~\citep[see][]{veill05}. This technique has the  inconvenient 
of lacking 
information on the kinematics of the outflow. Alternatively, winds have also
been studied through detection in the spectra of a 
blueshifted component in some interstellar absorption
lines ~\citep[see for example][]{heckman00,rupke05a,martin06}. This
latter approach 
lacks spatial resolution, and consequently does not provide
information on the wind geometry. 
We have taken a new approach by using integral field spectroscopy,
which combines spectroscopic and imaging techniques. We are therefore able
to detect absorbing material throughout the field of view and to produce a map 
of the distribution and kinematics of the absorbers. 
In addition, we have been able to discriminate kinematically the 
ionized component of the outflowing material and to produce maps
for the distribution and kinematics of this hotter phase of the wind.
This approach provides a unique simultaneous global view of 
these two phases of a galactic wind.
\section{Observations and Data Analysis}
\label{sectobs}
The data analyzed in this paper were obtained on 2002 March 16 at the Observatorio del Roque de los
Muchachos on La Palma with the fibre system INTEGRAL~\citep{arribas98}
at the William Herschel Telescope. The observational setup and 
basic data reduction are detailed in~\cite{afri07}. We only
mention here the aspects more relevant to the present work.
The observed spectral range (5600-6850 $\mathrm \AA$) contains a few strong interstellar 
emission lines ($\mathrm H\alpha$, [NII]$\lambda\lambda6548, 6584$, 
[SII]$\lambda\lambda6716, 6731$) as well as several stellar absorption lines (NaD doublet
and weaker, mostly blended FeI and CaI lines). 
%

A first inspection of the spectra in the fibres with the highest S/N reveals an 
unusual ratio of the NaD absorption doublet with respect to the other
stellar lines, with the NaD lines being deeper than expected 
~\citep[see Fig. 1 in][]{afri07}.
The most straightforward explanation for this is contamination by interstellar
NaD absorption. 
We therefore proceed to separate the stellar and interstellar components in
the fibre spectra. 
We use a synthetic stellar population of 1 Gyr with solar metallicity to
subtract the stellar component of the spectra ~\citep[see][for a detailed explanation]{afri07}. 
This synthetic spectrum is used as a template to fit the
spectrum of each fibre. The free parameters in the fit are the velocity, velocity 
dispersion and (global) line strength. It is important to remove from the fitting
the wavelength range containing  the NaD doublet, as it is likely to be
contaminated with interstellar absorption. The result of this procedure
reproduces very well the stellar features in the spectra 
except for the NaD lines. The stellar template fits the redmost part of the NaD lines,
leaving a strong interstellar residual which is clearly blueshifted with respect to
the stellar component~\citep[see Fig. 1 in][]{afri07}. The presence of blueshifted 
interstellar absorption lines
is an unambiguous signature of outflowing material~\citep[see for example][and references therein]{rupke05a}. 
After subtraction of the stellar component, a spectrum of the
interstellar material remains (with both, emission and absorption lines). 
We analyze these interstellar spectra throughout the rest of this paper.
\section{Morphology and Kinematics of the Neutral Na I Phase}
\label{sectmorna}
Most works on the study of outflows or galactic winds by means of the NaD absorption~\citep[e.g.][]{heckman00,rupke02, schwartz04,martin05, martin06,rupke05a} do either select samples of objects where
the NaD doublet is dominated by the interstellar component, or try to make
a global rough correction for the stellar contribution. We present here a spatially resolved separation of these two components. With this aim, we apply the analysis 
described in section \ref{sectobs} to the spectra of the 2D collection obtained from the 
central region of M100, deriving  
full 2D maps of the distribution and kinematics of the absorbing material. 
A first result of this study is that regions where the NaD is dominated
by the interstellar component coexist in the same galaxy with regions where the stellar
component is dominant and there is very little, if any, cold interstellar absorbing gas. We would like to point out at this point that 2D spectroscopy has been
critical in detecting
the wind thanks to its ability to isolate the locations
where interstellar absorption is relevant. Indeed, the wind might
have been unnoticed in a spectrum taken in a large aperture where stars
dominate the NaD absorption.

\begin{figure}
   \centering
   \includegraphics[angle=0,width=7.8cm, clip=true]{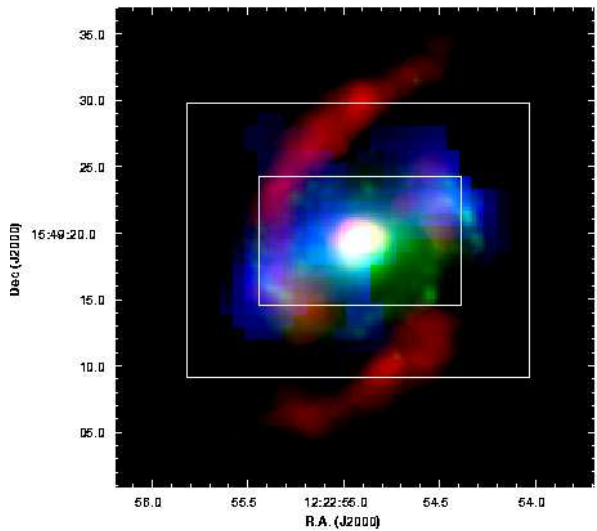}
   \caption{Composite RGB image of the equivalent width of the interstellar
NaD doublet (blue), K band image (green) and CO intensity map (red). Intensities are arbitrarily scaled. The FOV of SB3 (outer box corresponding to size of Figs. 2 and 3) and of SB2 (inner box corresponding to size of Figs. 6 and 7) have been marked.}\label{nadew}
\end{figure}

\begin{figure}
   \centering
   \includegraphics[angle=0,width=7.8cm, clip=true]{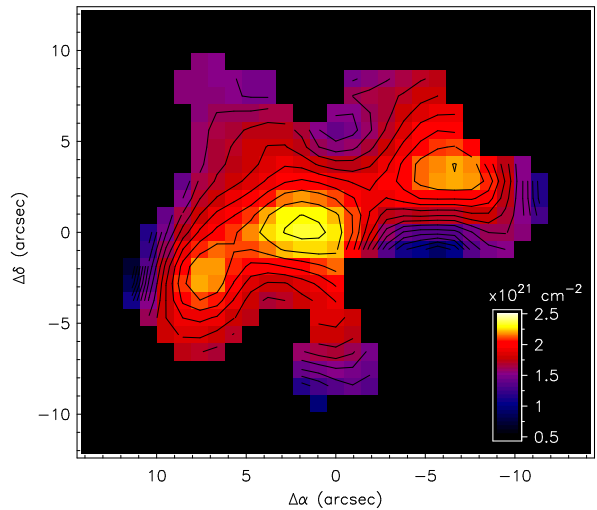}
   \caption{Map of hydrogen column density determined from the NaI absorption. Contours are spaced every 10$^{20}$ cm$^{-2}$.}\label{colhi}
\end{figure}

In Fig. \ref{nadew} we present the equivalent width
map for the interstellar NaD doublet, which ranges between 800 and 2100 $
{\rm m\AA}$ throughout
the field of view (lower values of the EW have been masked out).
The distribution of NaI in the map is 
clearly 
elongated along the bar position angle~\citep{knapen95} with a central maximum and two knots.
The central maximum coincides with the nuclear starburst and 
the location of the two knots matches that of the NIR knots corresponding
to active star forming regions. These knots are 
also related to two strong concentrations of absorbing material at 
these locations that are clearly seen in the CO map~\citep[from][]{saka95}.
With the exception of these knots, 
most of the gas responsible for the NaD absorption is concentrated in the 
region interior to the star formation ring. 
The determination of the column density of NaI from the spectra is, in general,  not 
straightforward ~\citep[see][and references therein]{martin06}. In the case under study, 
the situation is worsened because the stellar
component must be removed, which can seriously affect the shape of the final spectra.
Thus, the fitting of optical depth and covering factor as performed by~\cite{rupke05a} might not be fully reliable for all spectra (particularly but not exclusively for those with relatively low S/N ratio).
The equivalent width is much less affected by the
above-mentioned effects and is therefore a more {\bf reliable} estimator of the
column density. We then estimate the column density of NaI from 
the equivalent width of the NaD doublet using a curve of growth technique.
 We have taken a hybrid approach for this purpose. 
In first place, we use a direct fitting as in~\cite{rupke05a} on a 
selected subset of
high S/N spectra to estimate some average parameters  of the redder, $\lambda\lambda {\rm 5896}$, line: covering fraction, $C_f=0.17\pm 0.05$, Doppler parameter, $b=119\pm 35 {\rm km/s}$, 
and optical depth, $\tau_0=0.7\pm 0.3$.
In a second step, we use these average values and the curve of growth to find the 
following relation 
between the NaI column density and the equivalent width
of the NaD doublet:
${\rm N(NaI)\approx EW\times 2.88\times 10^{13} cm^{-2} }$.
Finally, to convert N(NaI) to hydrogen column density, we adopt the relation:
${\rm \log[N(NaI)]=\beta\log[N(H)]+\gamma}$ 
with $\beta=2.11$ and $\gamma=-31.3$ from~\cite{stokes78} as~\cite{rupke02}.
The resulting map of hydrogen column density is shown in Fig. \ref{colhi}.
From this map we have calculated the total hydrogen mass in the wind to be
${\rm \sim 2.4\times 10^6 M_\odot}$.
\begin{figure}
   \centering
   \includegraphics[angle=0,width=7.8cm, clip=true]{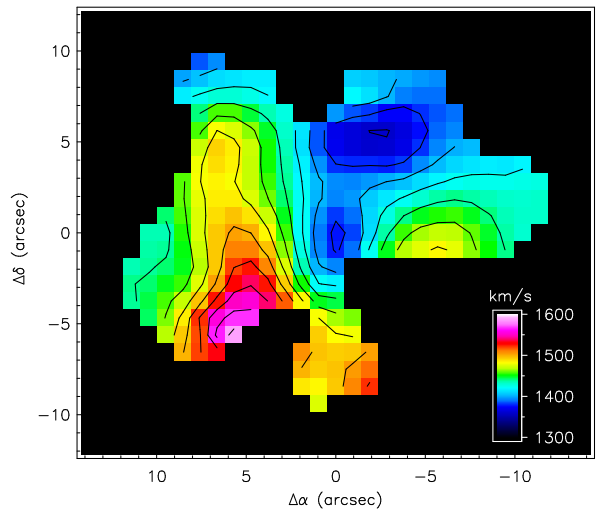}
   \caption{Velocity field of the absorbing NaI gas. Contours are spaced every 20 km/s.}\label{nadvel}  
\end{figure}
 The velocity at each fibre is calculated by fitting two gaussians to the NaD doublet in the (stellar removed) 
spectra. Errors in the velocity determination
range from 20 km/s to 70 km/s, with the highest errors occurring at
the locations with the lowest EW.
The velocity map of the absorbing gas (Fig. \ref{nadvel}) presents a rotation pattern 
with a major
kinematic axis orientation and velocity amplitude that clearly resembles that of the 
ionized gas in the disk~\citep[see][]{afri07}. However, this rotation pattern is globally blueshifted 
with respect to the underlying ionized gas in the galactic disk  
(${\rm v_{sys}=}$ 1567 km/s) by $\sim$ 115 km/s. 
The detection of this 
rotation pattern in the absorbing gas
indicates that this neutral phase is very likely disk material
that has been dragged upwards by the galactic wind, but that still keeps
most of its angular momentum.
It is probably material near the galactic disk, on
the interface between the fast, hot gas filling the cavity, and the underlying quiescent gas in the galactic disk. 

In order to measure the motion of the absorbing gas with respect to the 
underlying galaxy,
we have subtracted the velocity field of the ionized gas in the disk ~\citep[see][]{afri07} from the 
NaD velocity map.
The resulting  projected velocities range from
$\sim$ -210 km/s to roughly -55 km/s with an average of -115 km/s. 
\section{Morphology, Kinematics and Physical Conditions of the Ionized Gas Phase}
The distribution and kinematics of the ionized gas in the circumnuclear 
region of M100 has been extensively studied
by several authors~\citep[see for example][and references therein]{afri07}. However, 
after our finding of a strongly blueshifted component in the 
neutral gas traced
by the NaD lines, it is reasonable to wonder whether this outflowing gas 
has an ionized counterpart, and if this is the case, whether we can detect it. 
This detection has a  major difficulty with respect to the neutral gas, namely, 
the fact that we see the outflowing material in front of the bright background 
galaxy. Fortunately, integral field spectroscopic data provides kinematical 
information, and we may try to kinematically
discriminate these two ionized components ~\citep[see for example][]{arri93, arri94}.

\begin{figure}
   \centering
   \includegraphics[angle=0,width=8.5cm, height=3.8cm, clip=true]{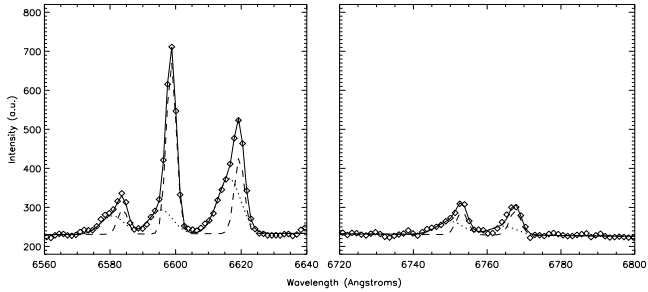}
   \caption{Plot of fibre \#89 where the detection of a double kinematical
component can be seen. Long and short dashed lines show the best fit to the 
disk and wind material respectively. Continous line represents the total fit.}
\label{twocompspec}
\end{figure}

Visual inspection shows indeed a blue shoulder in the emission
lines in many spectra taken with the SB3 and SB2 bundles (see, for instance, Fig. \ref{twocompspec}). Unfortunately, the spectral resolution
with the SB3 fibre bundle is not high enough (4.8 $\mathrm\AA$), and
this blue shoulder is only visible in a few fibres. The SB2 bundle
offers a much better (2.8 $\mathrm\AA$) spectral resolution, and the
line asymmetry is visible in many more fibres. We therefore perform this
kinematical discrimination in the SB2 bundle only.
In order to study this blueward component we try to fit two kinematical components to the emission lines of each spectra. To do this we 
simultaneously fit two sets of five spectral lines ($\mathrm H\alpha$, 
[NII]$\lambda\lambda6548, 6584$, [SII]$\lambda\lambda6716, 6731$). 
We assume that, for each set, all these five lines have
the same velocity and velocity dispersion. The theoretical 3:1 line ratio
is fixed for the [NII] doublet. Fig. \ref{twocompspec} shows the fit to a spectrum where the double component is clearly
detected. A kinematical double component is said to be detected only if the 
five spectral lines are detected in the two sets and the
reduced $\chi^2$ of the fit shows a significant improvement of at least 50\% with respect to the fit to a single kinematical component.  As a final check, each fit is inspected visually and
in case of doubt a single component is kept. 
When this procedure is finished, there is still some ambiguity at some positions 
on which of the
two components belongs to the background galaxy and which to the outflow. 
Continuity in the produced maps (particularly in the velocity field of the 
background galaxy), and the higher velocity dispersion of the secondary 
component are used to discriminate between the two.
\begin{figure}
   \centering
   \includegraphics[angle=0,width=7.8cm, clip=true]{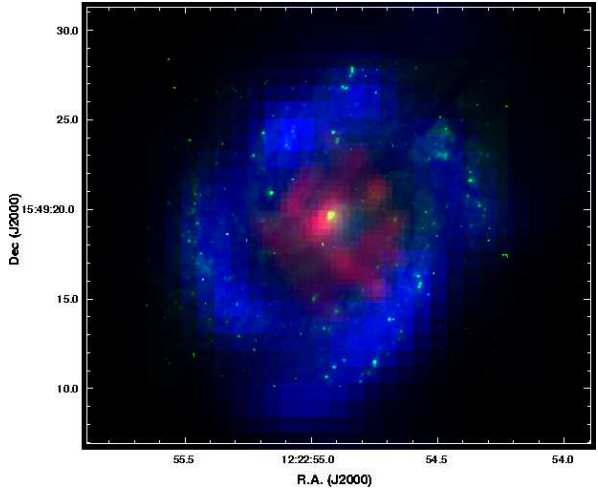}
   \caption{Composite RGB image of ${\rm H\alpha}$ intensity of the wind component (red), HST/ACS F555W (green) and ${\rm H\alpha}$ intensity of the underlying galaxy disk (blue). Intensities are arbitrarily scaled.}\label{haint2}
\end{figure}
\begin{figure}
   \centering
   \includegraphics[angle=0,width=7.8cm, clip=true]{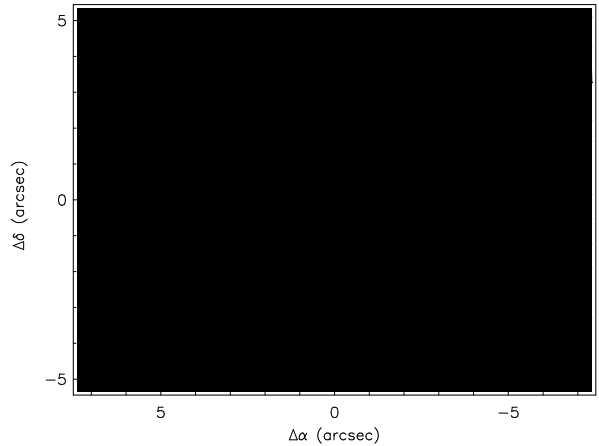}
   \caption{Velocity field of the ionized component of the wind with respect
to the underlying galaxy. Contours show the distribution of galactic H$\alpha$.}\label{havel2}
\end{figure}
In Figs.~\ref{haint2}, \ref{havel2} we present the intensity and velocity maps corresponding to the wind component of the ionized gas phase. The spatial distribution
of the ionized gas wind component is concentrated around the nuclear starburst
(although somewhat elongated along the major axis) 
with a bright ridge that extends towards the 
SW and a few weaker patches. The central and SW ridge may be enhanced 
emission from the walls of the expanding shell seen in projection.
The ionized component of the wind fills the region 
interior to the ring. If there is indeed an expanding shell in this region
it could be possible that the star forming ring is somehow related to the wind.

The velocity field of the ionized gas shows no rotation pattern. Nevertheless, the 
velocity field of the underlying ionized gas in the galaxy disk has been 
subtracted to view the motion of the wind with respect to it. 
It is blueshifted in most locations if we make exception of a 
redshifted region in the N. The maximum blueshifted velocity is located in the
 nucleus and
on a ridge extending from the S to the NW, which can be interpreted in terms 
of the projection of the walls of an outflowing cone or shell.
 These results are compatible with a bi-polar outflow
(in which case we could identify the region of receding velocities 
with the dimmed wind on the far side of the galaxy). 

\begin{figure*}
   \centering
   \includegraphics[angle=0,width=14cm, clip=true]{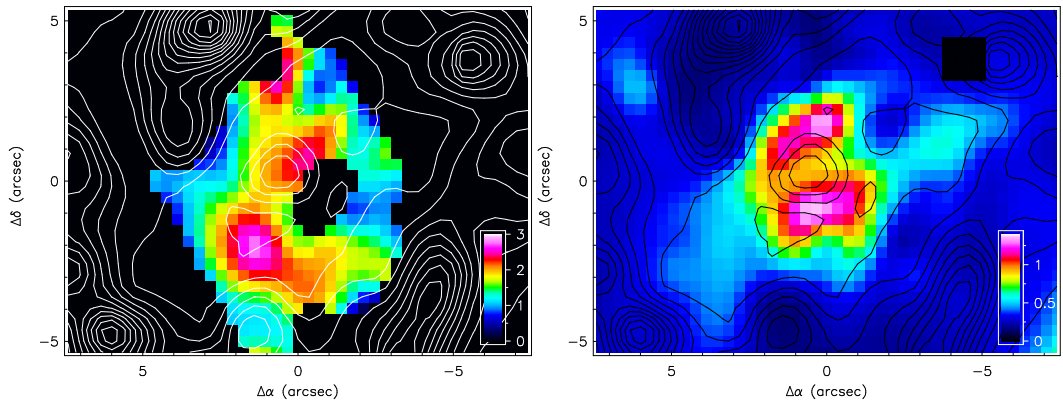}
   \caption{[NII]/H$\alpha$ line ratio for the wind (left) and galactic (right) components. Contours show intensity of galactic H$\alpha$ emission.}\label{liner}
\end{figure*}

The wind component has [NII]/H$\alpha$ ratios in the 1.2 to 2.7 range 
 and [SII]/H$\alpha$ in the 0.4 to 1.5 range, typical of shock ionization. The galactic component shows lower values, with [NII]/H$\alpha$ of 0.9 for the nucleus, and $1.1<\rm [NII]/H\alpha<1.4$ 
 ($0.3<\rm [SII]/H\alpha<0.7$) for the region inside the star forming
ring and of about 0.2 in this ring (see Fig. \ref{liner}). 
This indicates that even for the galactic
component, shock ionization is not negligible 
in the nucleus, and is even more important in the region inside the star 
forming ring. This region also shows the highest velocity dispersion 
(up to 100 km/s) of the region. 
Whether a hidden wind component sharing the disc kinematics is responsible
for the high line ratio and velocity dispersion in the galactic component 
in this region around the nucleus,
remains unclear. 
On the contrary, the ionization in the stellar formation ring appears not to be 
directly related to the wind.
The velocity dispersion of the ionized gas in the wind is in the range
between 150 and 230 km/s, while for the galactic component this value is much
lower, ranging between 50 and 100 km/s.  
\section{Discussion}
 In previous sections we have reported the discovery of an important wind
in the central region of M100. This is certainly an unexpected finding,
particularly because its extent 
($\approx {\rm 1 kpc}$) and velocity
($\approx {\rm 200 km/s}$) are quite large for
a galaxy with a moderate nuclear and/or star forming activity. 
In fact, the velocity of this wind is comparable to the values found by~\cite{martin06} for ULIRGs.
We have estimated the total mass of gas in the wind to be of about 
${\rm 2.4\times 10^6 M_\odot}$. Although this value is about an order of 
magnitude lower than
the values found by~\cite{martin06}, we must keep in mind that ULIRGs 
are forming stars one hundred times faster. This comparison makes
the wind in M100 even more spectacular and surprising.

Although the geometry of the wind is uncertain,
our findings indicate that the galactic wind is most likely driven uniquely by the 
nuclear starburst. The presence of rotation in the kinematics of the 
absorbing neutral gas indicates that it is likely disk ISM near the plane that  has been entrained by the
wind at the interface with the hot expanding shell. 
 The fact that we have not detected a rotation pattern in the 
wind component of the ionized gas, together with the high [NII]/H$\alpha$ 
and [SII]/H$\alpha$ ratios, indicates that this is shocked
gas at the top and/or walls of the expanding shell.
This scenario is also fully compatible with the irregular geometry of 
the extended continuum radio emission found by ~\cite{filho00}, and the diffuse
X ray emission reported by ~\cite{immler98}.
This new scenario also opens the possibility that the nuclear star forming ring
in M100 is related to this wind, either by direct compression of the ISM
at the edges of the shell, or by material of the wind falling down to the disk.
In fact, \cite{ryder01} found evidence for sequentially triggered star formation
in the ring of M100, compatible with this scenario.

Taking into account that we find projected 
velocities of up to $\sim$ 210 km/s and 
velocity dispersions of up to 220 km/s for
the ionized gas, there is
a non negligible fraction of the gas (with 
${\rm v_{wind}+2\sigma > v_{esc} \approx \sqrt{2}v_c(1+\ln (r_{max}/r))^{1/2}\sim 500 km/s}$) 
which might be able to escape the gravitational potential
of the galaxy (taken from a singular isothermal sphere as in ~\cite{rupke02}). 
Although this rough estimate should be taken with caution, it
indicates that winds of this type in the abundant galaxies with low nuclear activity 
may be very
important in polluting the IGM with dust and metals. The relevant question 
then is: how representative is the wind in M100 of similar phenomena in 
other normal, relatively unactive 
galaxies?. This question can only be answered by observing a
significant sample of low to moderate activity galaxies with the techniques used in this paper.
\section*{Acknowledgments}
This paper has been supported by the ``Secretar\'{\i}a de
Estado de Pol\'{\i}tica Cient\'{\i}fica y Tecnol\'ogica'' (AYA2004-08251-C02-02,ESP2004-06870-C02-02,AYA2006-02358).
This research has made use of NED
which
is operated by the JPL, Caltech, under contract with NASA. We have used observations made with the NASA/ESA HST, obtained from the data archive at the STScI.
J.J.V. acknowledges support from the Consejer\'{\i}a de
Educaci\'on y Ciencia de la Junta de Andaluc\'{\i}a.
A.C.M. acknowledges the support from the Ministerio
de Educaci\'on y Ciencia.
We also acknowledge the support of the RTN Euro3D: "Promoting 3D
spectroscopy in
Europe". We are grateful to the anonymous referee for valuable comments that
have helped to improve the quality of the paper.


\begin{thebibliography}{99}
\bibitem[\protect\citeauthoryear{Arribas \& Mediavilla}{1993}]{arri93} Arribas S., Mediavilla E. 1993, ApJ, 410, 552

\bibitem[\protect\citeauthoryear{Arribas \& Mediavilla}{1994}]{arri94} Arribas S., Mediavilla E. 1994, ApJ, 437, 149

\bibitem[\protect\citeauthoryear{Arribas et al.}{1998}]{arribas98} Arribas S. et al. 1998. Proc. SPIE, Vol. 3355, p.821. Optical Astronomical Instrumentation. D'Odorico S., Ed.

\bibitem[\protect\citeauthoryear{Bingham et al.}{1994}]{bingham94}Bingham R. G., Gellatly D. W., Jenkins C. R., Worswick S. P., 1994. Proc. SPIE, Vol. 2198, p.56. Instrumentation in Astronomy VIII. Crawford D. L \& Craine E. R. Eds.

\bibitem[\protect\citeauthoryear{Castillo-Morales et al.}{2007}]{afri07} Castillo-Morales A. Jim\'enez-Vicente J., Mediavilla E., Battaner E. 2007, MNRAS in press


\bibitem[\protect\citeauthoryear{Filho et al.}{2000}]{filho00} Filho M. E., Barthel P. D., Ho L. C. 2000, ApJS, 129, 93

\bibitem[\protect\citeauthoryear{Heed no recibe paquetesckman et al.}{2000}]{heckman00} Heckman T. M., Lehnert M. D., Strickland D. K., Armus L. 2000, ApJS, 129, 493

\bibitem[\protect\citeauthoryear{Ho et al.}{1997}]{ho97a} Ho L. C., Filippenko A. V., Sargent W. L. W. 1997, ApJS, 112, 315


\bibitem[\protect\citeauthoryear{Immler et al.}{1998}]{immler98} Immler S., Pietsch W., Aschenbach B. 1998, A\&A, 331, 601


\bibitem[\protect\citeauthoryear{Knapen et al.}{1995}]{knapen95} Knapen J. H., Beckman J. E., Shlosman I., Peletier R. F. Heller C.H., de Jong R. S. 1995, ApJ, 443, L73


\bibitem[\protect\citeauthoryear{Martin}{2005}]{martin05} Martin C. L. 2005, ApJ, 621, 227

\bibitem[\protect\citeauthoryear{Martin}{2006}]{martin06} Martin C. L. 2006, ApJ, 647, 222

\bibitem[\protect\citeauthoryear{Rupke et al.}{2002}]{rupke02} Rupke D. S., Veilleux, S., Sanders, D. B. 2002, ApJ, 570, 588

\bibitem[\protect\citeauthoryear{Rupke et al.}{2005}]{rupke05a}    Rupke D. S., Veilleux, S., Sanders, D. B. 2005, ApJS, 160, 87

\bibitem[\protect\citeauthoryear{Ryder et al.}{2001}]{ryder01} Ryder S. D., Knapen J. K., Takamiya M. 2001, MNRAS, 323, 663

\bibitem[\protect\citeauthoryear{Sanders et al.}{2003}]{sanders03} Sanders D. B., Mazzarella J. M., Kim, D.-C., Surace J. A., Soifer B. T. 2003, AJ, 126, 1607

\bibitem[\protect\citeauthoryear{Sakamoto et al.}{1995}]{saka95}
Sakamoto K., Okumura S., Minezaki T., Kobayashi Y., Wada, K. 1995, AJ, 110, 2075

\bibitem[\protect\citeauthoryear{Schwartz \& Martin}{2004}]{schwartz04}    Schwartz C. M., Martin, C. L. 2004, ApJ, 610, 201

\bibitem[\protect\citeauthoryear{Stokes et al.}{1978}]{stokes78} Stokes G. M. 1978, ApJS, 36, 115   
\bibitem[\protect\citeauthoryear{Veilleux et al.}{2005}]{veill05} Veilleux S., Cecil G., Bland-Hawthorn J. 2005, ARAA, 43, 769

\end{thebibliography}
\end{document}